\documentclass[traditabstract]{aa} 

\usepackage[dvips]{graphicx}          
\usepackage[comma,authoryear]{natbib} 
\bibpunct{(}{)}{,}{a}{}{,}            

\begin{document}

   \title{Bridges between helioseismological and asteroseismological inference}


   \author{D.O. Gough\inst{1,2}
          }

   \institute{Institute of Astronomy, University of Cambridge,
              Cambridge CB3 0HA, UK,
              \email{douglas@ast.cam.ac.uk}
              \and
              Department of Applied Mathematics and Theoretical Physics, 
              University of Cambridge CB3\,0WA, UK 
             }

  \abstract
{Exactly eighty years ago, a very young Yehudi Menuhin was invited by Bruno Walter to perform Beethoven's violin concerto with the Berlin Philharmonic Orchestra.  Walking through the streets of Berlin he was unsure of his way, and asked a passer-by how he could get to the Konzerthaus.  The man looked at him, looked down at the violin case that Yehudi was carrying, and said: `Practise, young man, practise'.  It was with such advice in their minds, I am sure, that Margarida and Michael have asked me to try to build bridges between helioseismology and asteroseismology.  Asteroseismology is new and fresh, and the young scientists who are entering the subject should be full of the expectation of the delights of discovery of untrodden ground.  Where should they tread?  They should be guided, perhaps, by our mature, well practised, experiences with the Sun.}


   \maketitle


With this anecdote in mind I recall some of the early developments in our soundings of the Sun.  I must attempt to confine attention to our inferences from low-degree modes alone, appreciating the difficulty of attempting to ignore the bias that was inevitably exerted by what we had learned from modes of high degree, knowledge of a kind that will not be obtainable from other stars.  Many of our early inferences were derived from calibrations of solar models, a necessary procedure because localized information (either in configuration space or in the space of more general enquiry) genuinely uncontaminated by global properties is almost impossible to obtain from modes of only low degree.  Moreover, accurate nonseismic information of the kind we have about the Sun, such as mass and radius, will not always be available.  We must bear in mind that seismic data alone can at best provide information about only the functional forms of sound speed and density with respect to fractional radius, which we might use, with the help of some idea about the value of the first adiabatic exponent $\gamma _1$ (which for stars composed of `ordinary' matter is close to 5/3 everywhere except in the outer layers where the effects of ionization of the abundant elements are important), to create a seismically calibrated model of the star.  Nonseismic information, be it in the form of observation or prejudice (otherwise known as prior information), is essential to `complete' the picture.

The first solar calibration from low-degree modes was carried out by J{\o}rgen Christensen-Dalsgaard and me for the purpose of estimating the helium abundance $Y$, a quantity of extreme importance at that time to the solar neutrino problem.  Two different fits to the data were better than the rest; they were about equally good (although not as good as we had hoped), so the calibration was indeterminate.  One had $Y$ above our expectation, the other below.  The reason that neither of the fits was as good as we had hoped was (partly) that the surface layers had been inadequately modelled, a matter which was not of principal interest {\it per se} -- after all, those layers are irrelevant to neutrino production -- but which impeded our diagnostic endeavours.  Actually, we knew which of the two fits to choose, from an earlier calibration of the convection zone with high-degree data, but that was outside the information space within which asteroseismologists can work, so I cannot discuss it here.  Nevertheless, I am reminded of some of the Bayesian discussions at this workshop.  Had we used Bayesian statistics with a neutrino-flux-based prior, we would have chosen the wrong fit.

That first calibration was extremely naive: it sought to find the model that best reproduces all the frequency data indiscriminately, based on a philosophy which, broadly speaking, is still in use today to provide `inversions' by (regularized) data fitting by least squares, a procedure commonly known as `regularized least squares'.

Today we seek frequency combinations that are signatures of some specific feature of the structure of the star that we might wish to investigate.  An example is to translate the inner phase of oscillation of some stellar model into a measurable frequency signature, in the manner proffered by Ian Roxburgh and Sergei Vorontsov.  Christensen-Dalsgaard has done that at this workshop for a Procyon-like model to reproduce the behaviour of the small frequency separations reported by Tim Bedding.  He is no doubt correct in claiming that the convective core provides the explanation of the curvature of Bedding's echelle plot, but I here play devil's advocate to warn that such behaviour could also be reproduced by appropriate -- some would rightly say contrived -- aspherical structure in the outer layers of the star.  We must invoke what one might misname a prior (actually a posterior constraint) to reject the latter.  The same comment applies to the measures of the different integral properties of the core (which do not require a specific stellar model for their determination) that were discussed here by G\"{u}nter Houdek in the context of the Sun.

I might point out that in the early days onlookers regarded our endeavours to be impossible.  I recall, for example, giving a lecture explaining how we expected to determine the solar $Y$ directly by measuring the effect of helium ionization on $\gamma _1$.   In the audience was Donald Lynden-Bell, who considered it impossible; he wagered that even after ten years' work we will not have succeeded.  He was right.  But not for the right reason.  We had actually achieved the impossible: in fact, we had measured the spatial variation of $\gamma _1$ so well that we knew that it contradicted all known equations of state, and therefore we could not trust those equations to convert the $\gamma _1$ variation into a reliable value of the helium abundance.

One might ask how it came about that, once the initial broad results of helioseismology had been established, the subject advanced so much further.  The reason was the promise to shed light on some big issues, principally the solar neutrino problem and the test of General Relativity via the precession of the orbit of Mercury, the latter requiring an evaluation of the oblateness of the Sun's gravitation field.  Substantial resources were made available to fulfill that promise, in the form of networks of ground-based observatories and the spacecraft SoHO.  Now that the most important questions have been answered, according to many scientists outside our subject there is little more of importance for helioseismology to do.  Consequently, funding has become more difficult, as several of us in this room have experienced.  Asteroseismologists should heed the warning that this situation presents.  We are riding on the backs of the planet hunters, and it might be wise to address some of the issues with which they are concerned.  To be sure we are acutely aware of the importance of estimating the radii of stars that planets might transit -- which of course requires supplementation by nonseismic information -- but in addition we could profitably consider other issues that concern the study of the formation of planets.  For example, it is mooted that chaotic orbits in newly formed planetary systems lead to some of the planets falling into their parent stars, contaminating the outer (convecting) layers of those stars with material rich in heavy elements.  Therefore enriched convection zones might be a characteristic of planet-hosting stars (even though this idea does not neatly explain all the existing spectroscopic data).  Can this hypothesis be tested seismologically?  If the accreted material were to remain confined to the convection zone, the oscillatory (with respect to frequency) contribution to the mode frequencies induced by the chemical discontinuity at the base of the zone could be sought -- but only if that discontinuity has not been destroyed by the fingering instability.  Unfortunately, the flux of material transported by fingers in stars is not known.  I have made a very crude estimate which suggests that the discontinuity is quickly washed out, and Sylvie Vauclair has come to a similar conclusion, but Pascale Garaud is currently carrying out a far superior investigation via (two-dimensional) numerical simulations, which appear to suggest that the discontinuity might be preserved for 10$^9$ years or so.  If she is right, and I hope she is, an important test will be available to us.

It behoves me to mention that we helioseismologists are dissatisfied with the state of our inferences beyond the seismically accessible aspects of the structure of the Sun.  Perhaps the most prominent problem at the moment is how to confront the low photospheric abundances reported by Martin Asplund and his colleagues of the principal opacity-producing chemical elements.  The crux of the problem can be stated quite simply:  We know the sound speed and density throughout the Sun from helioseismology.  If we now accept the nuclear reaction rates -- and I recommend that we do because the reaction cross-sections have long been studied in minute detail over decades in the quest for a resolution to the solar neutrino problem, and incorporating them into thermonuclear rates is adequately well understood (uncertainties in electron screening of fast nuclear encounters, for example, are too small to have a material effect on this argument) -- and if we adopt the generally accepted age of the Sun (together with the other assumptions of solar-evolution theory, often unstated in the literature), then one can make a good estimate of the abundance of hydrogen fuel, and hence, through the equation of state, infer the temperature.  We now have every variable in the equation of radiative transfer in the deep interior save the opacity, $\kappa$.  Hence we can infer a value of $\kappa (r)$ from that equation.  The problem is to reconcile that with Asplund's measurements.

Perhaps the most direct reconciliatory suggestion has been made by Joyce Guzik: that the Sun has been contaminated by material deficient in heavy elements so that the abundances in the photosphere are lower than those beneath the convection zone, a suggestion at odds with many of the experts in planet formation.  Provided that the contaminating material is not too rich in helium, the molecular-mass discontinuity at the base of the convection zone would be stable, and the abundance difference could no doubt be maintained.  However, the  oscillatory signature in the low-degree eigenfrequencies produced by the discontinuity (if there is little or no compensating $Y$ discontinuity) would be of a magnitude much greater than is observed.  This is evidence against the hypothesis, although the consequences of putative limited mixing that might reduce the amplitude of the seismic signature has not been investigated with care.

Another potential resolution was hinted at, perhaps inadvertently, by Christensen-Dalsgaard at this workshop, and is perhaps also suggested by the calibrations reported by Houdek, although I am quite sure that what I am about to say is too extreme (and inadequately digested) to be plausible to either Christensen-Dalsgaard or Houdek.  It is that the Sun has aged more than is generally accepted, corresponding to an evolutionary time of a few tenths gigayears.  I choose my words carefully to avoid contradicting theories of planet formation, all of which limit the age of the Sun to being no more than 10$^7$ years or so greater than that of the oldest meteorites.  So we cannot afford the time required to make the Sun older.  We must instead contemplate more rapid ageing.  If the Sun had initially been more massive than it is today, for example, and had lost mass on the main sequence, it would have evolved more rapidly.  Alternatively, had the Sun's core initially been deficient in hydrogen, it would seem to be older today;  when the Sun condensed from the nebula in which the planets were also forming, might it not have preferred to do so in the gravitational potential well of a nascent super-Jupiter?  It would have taken only a few Jupiter masses of heavy material to do the trick.  As I intimated just now, these suggestions are unlikely to be correct.  But I mention them to emphasize how important it is to consider possibilities that lie outside the domain sampled by typical MCMC and genetic algorithms, useful as those procedures are.

Finally, I must remind you that the road ahead is not going to be easy.  J\'{e}r\^{o}me Ballot, Vorontsov and especially Daniel Reese have given us a glimpse of some of the complexity that we are likely to have to face.  We need to have the optimism that we shall be able to extract useful information from the real and apparent chaos that we shall encounter, a degree of optimism that was admirably demonstrated by Markus Roth, who computed from an artificial solar meridional flow some twenty-five times faster than he expects to be present in the Sun a signal that looks like zero.  And still he says that he hopes to detect the real flow!  That demonstrates the kind of optimism that we need.  Unless one really tries to achieve the impossible, there can be little chance of actually achieving it.

\end{document}